\documentstyle[preprint,12pt,aps]{revtex}
\tightenlines
\begin{document}
\title{{\boldmath $W_R$} Effects on 
   {\boldmath $CP$} Asymmetries in 
   {\boldmath $B$} Meson Decays}
\author{Dennis Silverman}
\address{Department of Physics, \\ 
 University of California, Irvine \\
 Irvine, CA 92697-4575} 
\author{Herng Yao}
\address{ Department of Physics, National Taiwan Normal  
 University, \\ Taipei, Taiwan 117}
\date{\today}
\maketitle

\begin{abstract}
After we have limited elements of the right-handed CKM matrix to satisfy the
bounds for $CP$ violation $\epsilon_{K}$ in $K$ meson systems, the
right-handed charged current gauge boson $W_R$ is shown to
substantially affect $CP$ asymmetries in $B$ systems. A joint
$\chi^{2}$ analysis is applied to $B-{\bar B}$
mixing to constrain the right-handed CKM matrix
elements. In $(\sin{(2\alpha)}, \sin{(2\beta)})$, $(x_s,
\sin{(\gamma)})$, $(\rho, \eta)$, and $(x_s,\sin{(2\phi_s)})$ plots in 
the presence of the
$W_{R}$ boson, we find larger allowed experimental regions that 
can distinguish this model from the standard model.
\end{abstract}


\newpage

\section{Introduction}

Within the standard model (SM), the flavour non-diagonal couplings in
the weak charged-current interactions are described by the unitary
Cabibbo-Kobayashi-Maskawa (CKM) matrix\cite{cabibbo}.  The SM has been
considered as the complete description of the weak interactions.
However, it is widely believed that there must be physics beyond the
SM. The left-right symmetric model (LRSM) is one of the simplest
extensions in new physics.  Currently, $B$ factories 
at SLAC and KEK have started to measure the $CP$ violating
asymmetries in the decays of $B$ mesons and provide a test of the SM
explanation of $CP$ violation.  The goal of this paper is to examine
the possible effects of a right handed boson $W_{R}$ on the
determinations of $CP$ violating decay asymmetries.

\section{Left-Right Symmetric Models}

The $V-A$ structure of the weak charged currents was established after
the discovery of parity violation\cite{lee}.  This is manifested in
the standard model by having only the left-handed fermions transform
under the $SU(2)$ group.  It is then natural to ask whether or not the
right-handed fermions take part in charged-current weak interactions,
and if they do, with what strength .  Charged-current interactions for
the right-handed fermions can easily be introduced by extending the
gauge group\cite{pati}. The simplest example is the $SU(2)_{L} \times
SU(2)_{R} \times U(1)_{B-L} $ model, where the left-handed fermions
transform as doublets under $SU(2)_{L}$ and as singlet under
$SU(2)_{R}$, with the situation reversed for the right-handed
fermions\cite{mohapatra}. The addition of a new $SU(2)_R$ to the gauge
group implies the existence of three new weakly interacting gauge bosons: 
two are charged and one is neutral.

  The charged right-handed gauge bosons (denoted by $W_{R}^{\pm}$) and
a neutral gauge boson $Z_{2}$ acquire masses, which are proportional
to a vacuum expectation value, and which become much heavier than
those of the usual left-handed $W_{L}^{\pm}$ and $Z_{1}$
bosons. The charged current weak interactions can be written as
(suppressing the generation mixing)
\begin{eqnarray}
 {\it L} &=& {g \over {\sqrt 2}} ({\bar
    u}_{L}\gamma_{\mu}d_{L} + {\bar
    \nu}_{L}\gamma_{\mu}e_{L})W_{L}^{+} \cr 
& & + {g \over {\sqrt 2}} (
    {\bar u}_{R}\gamma_{\mu}d_{R} + {\bar
    \nu}_{R}\gamma_{\mu}e_{R})W_{R}^{+} + {\rm H.C.}, 
\end{eqnarray}
where the gauge coupling for left and right handed currents 
is assumed to have the same strength $g$, and a $W_{L}-W_{R}$ 
mixing term is neglected since it is highly suppressed by the experimental 
data\cite{langacker}. 
It is clear that for $m_{W_{L}} \ll m_{W_{R}}$, the charged current
weak interactions will appear almost maximally parity-violating at low
energies. Any deviation from the pure left-handed (or $V-A$) structure
of the charged weak current will constitute evidence for a
right-handed current and therefore a left-right symmetric structure of
weak interactions.

Within the $SU(2)_{L} \times SU(2)_{R} \times U(1)$ model, we denote
the left- and right- handed quark mixing matrices by $V^{L}$ and
$V^{R}$, respectively. The form of $V^{L}$ is parametrized
by\cite{wolfenstein}
\begin{equation}
V^{L} = 
\bordermatrix{
   & d                         & s                          & b         \cr    
u  & 1- {\lambda^2 \over 2 }   & \lambda &  A\lambda^{3}(\rho - i\eta)  \cr 
c  &-\lambda                   & 1 - { \lambda^{2} \over 2} & A\lambda^2\cr
t  &A\lambda^{3}(1-\rho-i\eta) &  -A\lambda^{2}             &1          \cr}.
\end{equation}
On the other hand, the choice of the manifest left-right 
symmetry, i.e., $V^R=V^L$, would yield a very strigent 
bound $m_{W_{R}} \ge 1.6$ TeV from the constraint of the kaon
mass difference\cite{beall}. The limits from $\Delta m_K$
for arbitary $V^R$ were first considered by Olness and Ebel\cite{olness} 
and followed by Langacker and Sankar\cite{langacker}.
They showed that the lower limit of the $W_R$ mass could
be reduced by taking either of the following two general 
forms of $V^R$

\begin{equation} 
V_{I}^{R}
 = \bordermatrix{  
   & d           & s & b             \cr
u  & 1           & 0 &     0         \cr
c  & 0  & ce^{i\xi}  & s e^{i\sigma} \cr
t  & 0  & se^{i\phi} & ce^{i\chi}    },
   ~~~~V_{II}^{R} = \bordermatrix{ 
  & d            & s & b            \cr
u & 0 & 1        &     0        \cr
c & ce^{i\xi}    & 0 & s e^{i\sigma}\cr
t & se^{i\phi}   & 0 & ce^{i\chi}   },
\end{equation}
where $s=\sin{\theta}$ and $c=\cos{\theta} ~(0 \le \theta \le 90^{o})$, 
along with the 
unitarity condition $\xi - \sigma = \phi - \chi + \pi$. The former
type will be called the case of st-coupling ($V_I^R$) and the latter 
that of dt-coupling ($V_{II}^R$) 
in the paper. There is the possibility of having an overall
phase factor $e^{i\omega}$ multiplying both matrices, but in all
relevant processes considered here, the $W_R$ is not mixed but reabsorbed,
so these phase factors are always cancelled by their complex conjugates
and are not independent variables for statistical purposes.  We therefore
take $\omega$ to be zero for simplicity.

A general parametrization of $V_R$ will involve six phases\cite{barenboim}.
There is not enough data to constrain this many variables.  
As shown below, the contributions to $\epsilon_K$ from $V^R$
from $t$ and/or $c$ quarks in the inner box could be
a thousand times the standard model, unless products of $V^R$ matrix 
elements were one part per thousand.  Since the error on $\epsilon$ is
only 10\% to 15\% from $B_K$, these matrix elements would have to be small
and finely tuned to contribute.  So it is more natural to assume that
they are essentially zero for our calculations.  This means that the
$c$ and $t$ elements of the $d$ or $s$ columns are taken to vanish, 
as in the cases of st- and dt-coupling, respectively, in equation (3).

For our purposes 
of showing the effects of LRSM in $B$ physics $CP$ violating experiments, we
use the above cases as starting points.  They will each be shown to bring
in only one independent right hand phase.

\section{{\boldmath $CP$} Violation in {\boldmath $K$} Meson Systems}

The CP violation parameter $\epsilon_{K}$ in $K$ decays, which is
proportional to the imaginary part of the box diagrams mediated by two
$W_{L}$, or two $W_{R}$ or a $W_{L}-W_{R}$ pair, is given as
$\epsilon_{K} \approx {\rm Im} {\langle  K^{0}| H(\Delta S = 2) |
 {\bar K}^{0} \rangle} / 
{\sqrt 2} \Delta m_{K} $ where $H(\Delta S =2) =H^{LL} + H^{RR} +
H^{LR}$ is the Hamiltonian from the box diagrams named above.
The $H^{LL}$ contribution gives\cite{buras}
\begin{equation}
\epsilon_{K} = { G_{F}^{2} f_{K}^{2} B_{K} m_{K} m_{W_{L}}^{2}  \over 
 12\sqrt{2} \pi^{2} \Delta m_{K} } 
[ \eta_{cc}S(x_{c})I_{cc} + \eta_{tt}S(x_{t})
 I_{tt} + 2\eta_{ct}S(x_{c}, x_{t})I_{ct} ],
\end{equation}
where $I_{ij} ={\rm Im}(V_{id}^{*}V_{is}V_{jd}^{*}V_{js}),$ 
and the Inami-Lim functions [7] are
\begin{equation}
S(x) = x \left[{1 \over 4} + {9 \over 4(1-x) } - 
{3 \over 2(1-x)^{2} } \right]
        -{3 \over 2} ({x \over 1-x })^{3} \ln{x}, 
\end{equation}
\begin{equation}
S(x_{c},x_{t}) = x_{c}
\left[\ln{x_{t} \over x_{c}} - { x_{t} \over 4(1-x_{t}) }
           (1+{ x_{t} \over 1-x_{t} }\ln{x_{t}})\right], 
\end{equation}
with $x_{i} = m_{i}^{2} / m_{W_{L}}^{2}. $ 
The factors $\eta_{cc} = 1.38, \eta_{tt} = 0.59$, and $\eta_{ct} = 0.47$
are QCD corrections\cite{herrlich}.

The two $W_R$ exchange part, $H^{RR}$, gives no contribution due to the factor 
$I_{ij}$ vanishing for both
cases of $V^{R}$ as shown in eq. $(3)$. 

The part from $W_L$ and $W_R$ exchange, $H^{LR}$, gives\cite{ecker}
\begin{eqnarray}
  H^{LR} &=&
{2G^{2}_{F} \over \pi^{2} } 
 m_{W_{L}}^{2} x_{LR} \sum_{i,j=u,c,t}\lambda_{i}^{LR}\lambda_{j}^{RL} 
({\bar d_{R}}s_{L})({\bar d_{L}}s_{R}) 
 { \sqrt{x_{i}x_{j}} \over 4 } \\ \nonumber
& & [(4\eta_{ij}^{(1)}+\eta_{ij}^{(2)}x_{i}x_{j} x_{LR})
I_{1}(x_{i},x_{j},x_{LR})
-(\eta_{ij}^{(3)}+\eta_{ij}^{(4)} x_{LR})I_{2}(x_{i},x_{j},x_{LR}) ], 
\end{eqnarray}
where
$\lambda ^{LR}_{i} =V^{L*}_{id}V^{R}_{is}$, 
the ratio of $W$ masses squared is $x_{LR} = (m_{W_{L}}/m_{W_{R}})^{2}$,
\begin{eqnarray}
I_{1}(x_{i},x_{j},x_{LR})
&=&{x_{i}\ln{x_{i}} \over (1-x_{i})(1-x_{i} x_{LR})
(x_{i}-x_{j})} +( i \leftrightarrow j) \\ \nonumber
& &-{x_{LR} \ln{x_{LR}} \over (1-x_{LR})
(1-x_{i} x_{LR})(1-x_{j} x_{LR})}, \\
 I_{2}(x_{i},x_{j}, x_{LR})
&=&{x_{i}^{2}\ln{x_{i}} \over (1-x_{i})(1-x_{i} x_{LR}) 
(x_{i}-x_{j})} +( i \leftrightarrow j) \\ \nonumber
& & -{ \ln{x_{LR}} \over (1-x_{LR})
(1-x_{i} x_{LR})(1-x_{j} x_{LR})},
\end{eqnarray}
and $\eta^{(1)-(4)}$ are the short-distance QCD correction factors, whose
explicit forms are given in Ref.~\cite{ecker}.
Their values are 
$(\eta_{cc}^{(1)},\eta_{ct}^{(1)},\eta_{tt}^{(1)})=(0.61,1.27,1.98)$, 
$(\eta_{cc}^{(2)},\eta_{ct}^{(2)},\eta_{tt}^{(2)})=(0.04,0.27,0.75)$, 
$(\eta_{cc}^{(3)},\eta_{ct}^{(3)},\eta_{tt}^{(3)})=(0.55,1.03,1.93)$, and
$(\eta_{cc}^{(4)},\eta_{ct}^{(4)},\eta_{tt}^{(4)})=(0.45,0.84,1.58)$
for $m_{W_{R}}=2.5$ TeV and $m_{t} = 175$ GeV at the scale $\mu=4.5$ 
GeV.

The contribution to $\epsilon_{K}$ from $H^{LR}$ only comes from the
following combinations of quark mixing elements surviving in
$\lambda_{i}^{LR}\lambda{j}^{RL}$: for the case of st-coupling
\begin{eqnarray}
 {\rm (cu~pair)} &:& \lambda^{2} c \sin{(- \xi)}, \\
 {\rm (tu~ pair)}&:&A\lambda^{4}[(1-\rho)\sin{(\phi)}+
\eta \cos{(\phi)}] ;
\end{eqnarray}
 and for the case of dt-coupling
\begin{eqnarray}
{\rm (uc~pair)}&:&(1- {\lambda^{2} \over 2 })^{2} c \sin{(-\xi)}, \\
{\rm (ut~pair)}&:&As\lambda^{2}(1-{\lambda^{2} \over 2})
\sin{(-\phi)}. 
\end{eqnarray}
Again, the contributions to $\epsilon_K$ would be of the same order
as in the SM in the case of st-coupling, or 10 to 100 times as large in 
that of dt-coupling, 
unless some of the parameters were very small or if a cancelation occured.  
So we will adjust the parameters in $V^{R}$ so
that no contribution to $\epsilon_{K}$ will come from $H^{LR}$. This
is accomplished by various conditions\cite{kurimoto} in the two cases.
For the case of st-coupling ($V_{II}^R$), the LRSM model needs effectively:
\begin{equation}
\sin{(\xi)} = 0 ~~~{\rm and}~~~~ 
\tan{(-\phi)} = {\eta \over (1-\rho)}. 
\end{equation}
From the geometry of the unitarity triangle in the $\rho,\eta$ plane, we
see that $\phi = -\beta$, where $\beta$ is the unitarity 
triangle angle at $\rho=1$.
In this case, using the unitarity relation also, we have remaining two 
$V_I^R$ variables to vary: $s$ and $\sigma$.
For the case of dt-coupling the LRSM model effectively needs
\begin{equation}
c=0~~~~{\rm and}~~~~\sin{(\phi)} = 0. 
\end{equation}
In this case we have only the variable $\sigma$ in $V_{II}^R$ to vary.  
There are other solutions to suppress $\epsilon_{K}$. However,
we use the above cases since they give the most significant effects on $CP$ 
violation
in $B$ decay by $W_{R}$.  In summary then, the right handed mixing matrices
have become:
\begin{equation} 
V_{I}^{R}
 = \bordermatrix{  
   & d  & s            & b                   \cr
u  & 1  & 0            &     0               \cr
c  & 0  & c            & s e^{i\sigma}       \cr
t  & 0  & se^{-i\beta} & -ce^{i(\sigma-\beta)}    },
   ~~~~V_{II}^{R} = \bordermatrix{ 
  & d   & s   & b                  \cr
u & 0   & 1   & 0                  \cr
c & 0   & 0   & e^{i\sigma}        \cr
t & 1   & 0   & 0   } .
\end{equation}
Just as in the left handed CKM matrix where there is one somewhat sizeable
parameter ($\lambda$) mixing the lightest generations,  
while the other matrix elements are rather small, 
a similar behavior is seen in the right hand matrix.  In the case of 
st-coupling ($V_I^R$), 
it is the heavier generations that mix more than the lightest one, and
in the case of dt-coupling ($V_{II}^R$) it is the light s and d quarks 
whose mixing elements switch roles. 
Once the disappearance of the LR contribution to $\epsilon_K$ is arranged,
which comes from the imaginary part of the $M_{12}^{LR}$ matrix element, we
also find that the contribution of the real part, which gives $\Delta m_K$,
is also very small:  there is no contribution in the case of dt-coupling, and 
in the case of st-coupling, 
the contributions are only from $(t,u)$ and $(c,u)$ intermediate quarks.  These
contributions have the same orders of $\lambda$ as in the SM, but are 
suppressed by $m_u/m_t$ and $m_u/m_c$, respectively.

Recently, some detailed analyses of $W_{R}$ effects on $\epsilon_{K}$ 
have been made\cite{barenboim} in the LRSM models in which the third 
generation was not dominant.    
The large $N_{C}$ expansion and chiral 
perturbation theory were applied to estimate the left-right hadronic matrix
elements $ \langle K^{0}|H^{LR}(\Delta S=2)|\bar{K^{0}} \rangle $ and their 
uncertainties. 
The models considered here have dominant third generation effects and are 
complementary to the other models.

What we find here is that even with the mixing matrices $V_R$ in both cases 
that avoid the $\epsilon_K$ and $\Delta m_K$ constraints, 
the other experiments
still give a lower bound of
$m_{W_R} \ge 1.3$ TeV in the case of dt-coupling, at 95\% CL.  In this case, 
even though $V^R_{ud}=0$ in $W_R$ production, the experimental D0 limit from 
Fermilab\cite{abachi} is $m_{W_R} \ge 505$ GeV at 95\% CL (see Fig. 3
of ref. 13.).  In the case of st-coupling, with
$V^R_{ud}=1$, the Fermilab limit is $m_{W_R} \ge 720$ GeV, and we
present analyses for $m_{W_R} \ge 1$ TeV. With only the set of experiments
we have considered, the case of st-coupling does have solutions down to
$m_{W_R} \ge 0.25$ TeV.

\section{{\boldmath $B^{0}-{\bar B}^{0}$} Mixing} 

The mixing parameter $x_{q}$ in the $B^0_q-{\bar B^0_q}$ system is
defined by
\begin{equation}
x_{q} \equiv { (\Delta m)_{B_{q}} \over \Gamma } =
2\tau_{B_{q}}|M_{12}|,
\end{equation}
where $q = d$ or $s$, and $M_{12}$ is the dispersive part of the
 mixing matrix element, $i.e.,$ $M_{12} -{i \over 2}\Gamma_{12} =
 \langle B^{0}| {\it H} (\Delta B = 2) | \bar{B}^{0} \rangle .$ In the 
standard model, the mixing is explained by the dominant contribution of the
 two $t$-quark box diagrams.  In the LRSM, $M_{12}$ contains three
 terms
\begin{equation}
M_{12} = M_{12}^{LL} + M_{12}^{RR} + M_{12}^{LR},
\end{equation}
corresponding to the contributions from box diagrams in which two $W_L$,
two $W_R$ and a $W_{L}-W_{R}$ pair are exchanged.
The standard model matrix element $M_{12}^{LL}$ is
\begin{equation}
M_{12}^{LL} = {G^{2}_{F} \over 12\pi^{2} } m_{B}m_{W_{L}}^{2}(f_{B}^{2}B_{B})
    \eta_{tt}S(x_{t}) (V_{tq}^{L*}V_{tb}^{L})^{2} 
\end{equation}
where $S(x_{t})$ is defined in eq. $(5)$ and $\eta_{tt} = 0.59$ is the QCD 
correction factor.  The evaluation 
of the hadronically uncertain $f_{B}^{2}B_{B}$ has been the subject of much
work, which is summarized in Ref.~\cite{ali}. We will use
\begin{equation}
f_{B_{d}}B_{B_{d}}^{1/2}=210 \pm 40~ {\rm MeV~ and}~~
  f_{B_{s}}B_{B_{s}}^{1/2}=230 \pm 45~ {\rm MeV}  
\end{equation}
from the scaling law and recent lattice calculations.

The element $M_{12}^{RR}$ is given by
\begin{equation}
M_{12}^{RR} = {G^{2}_{F} \over 12\pi^{2} } m_{B}m_{W_{L}}^{2}(f_{B}^{2}B_{B})
    \eta_{tt}S(x_{t}) x_{LR}^{2} (V_{tq}^{R*}V_{tb}^{R})^{2}.  
\end{equation}
It disappears in $B_{d}-{\bar B_{d}}$ mixing due to either $V_{td}^{R} = 0$ or
$V^{R}_{tb} = 0$ for both cases in $V^{R}$, but it has a contribution
for the case of st-coupling in $B_{s}-{\bar B_{s}}$ mixing due to the non-zero
values of $V_{ts}^{R}$ and $V_{tb}^{R}$.

The matrix element $M_{12}^{LR}$ is 
\begin{eqnarray}
 M_{12}^{LR} &=&
{G^{2}_{F} \over 2\pi^{2} } m_{B}(f_{B}^{2}B_{B})({m_{B} \over
m_{b}})^{2} m_{W_{L}}^{2} x_{LR}
\sum_{i,j=u,c,t}\lambda_{i}^{LR}\lambda_{j}^{RL} \\  
& &\left[ { \sqrt{x_{i}x_{j}} \over 4 } 
[(4\eta_{ij}^{(1)}+\eta_{ij}^{(2)}x_{i}x_{j}x_{LR})I_{1}(x_{i},x_{j},x_{LR})
-(\eta_{ij}^{(3)}+\eta_{ij}^{(4)}x_{LR})I_{2}(x_{i},x_{j},x_{LR})] \right], 
\nonumber 
\end{eqnarray}
where $\lambda ^{LR}_{i} =V^{L*}_{iq}V^{R}_{ib}$,
$\lambda ^{RL}_{j} =V^{R*}_{jq}V^{L}_{jb}$, $\eta^{(1)-(4)}$,
and $I_{1}$ and $I_{2}$ are defined in Eqs. (7), (8) and (9).
In order to obtain this formula, 
the following ratio of matrix elements of quark operators\cite{ecker} 
has been applied
\begin{equation}
{ { \langle B^{0}|(\bar{d}_{R}b_{L})(\bar{d}_{L}b_{R})|\bar{B}^{0} \rangle} 
\over {\langle B^{0}|(\bar{d_L}\gamma_{\mu}b_L)^{2}|\bar{B}^{0} \rangle}  } =
{ { \langle B^{0}|\bar{d}_{R}b_{L}|0\rangle 
\langle0|\bar{d}_{L}b_{R}|\bar{B}^{o} \rangle} \over 
{\langle  
B^{0}|\bar{d_L}\gamma_{\mu}b_L|0\rangle 
\langle0|\bar{d_L}\gamma_{\mu}b_L|\bar{B 
}^{0} \rangle}}  = {3 \over 4} { B'_{B} \over B_{B}} 
( {m_{B} \over m_{b}} )^{2}, 
\end{equation}
where the bag factors $B'_{B}$ and $B_{B}$ encompass all 
possible deviations from the vacuum saturation approximation. 
$B'_{B}$ can be treated as approximately equal to $B_{B}$. Their  slight
difference is irrelevant compared to other uncertainties.

The contributions of the nine different combinations within Eq. (22)
are dominated by $(t,t),(t,c), (c,t)$ and $(u,t)$ pairs, for which the
values of the large square bracket at $m_{W_{R}} = 1$ TeV are $11.9$,
$4.6\times10^{-2}$, $5.0\times 10^{-2}$ and $0.80 \times 10^{-2}$,
respectively, the ratios mainly due to the quark mass factors $\sqrt
{x_{i}x_{j}}$.  All of the remaining terms are less than $10^{-3}$.

\subsection {{\boldmath $B_{d}-{\bar B}_{d}$} mixing}
\subsubsection{st-coupling}
In the matrix element $M_{12}^{LR}$ of Eq. (22) only two terms from
$(c,u)$ and $(t,u)$ pairs will survive in the case of st-coupling because of 
the 
factor $\lambda^{LR}_{i}\lambda^{RL}_{j}$.  One finds that $M^{LR}_{12} \ll
M_{12}^{LL}$ by four orders of magnitude, no matter what the mass
value $m_{W_{R}}$ is. Therefore, we may neglect the $W_{R}$
contribution to $B_d-{\bar B}_d$ mixing in this case.
\subsubsection{dt-coupling}
On the other hand, there is only one non-vanishing term from the
$(c,t)$ pair in $M^{LR}_{12}$ in the case of dt-coupling. One obtains 
$M_{12}^{LR} \sim M^{LL}_{12}$ if $m_{W_{R}} = 2.5$ TeV,
$M_{12}^{LR} <
M^{LL}_{12}$ if $m_{W_{R}} = 5$ TeV,
and $M_{12}^{LR} \le 10^{-2}  
M^{LL}_{12}$ if $m_{W_{R}} = 10$ TeV.  The effect from $W_{R}$ in
$B_d-{\bar B}_d$ mixing appears in this case.
        
\subsection{{\boldmath $B_{s}-{\bar B}_{s}$} mixing}
\subsubsection{st-coupling}
The effect from two $W_{R}$ exchanges appears here. $M_{12}^{RR} \ll
M^{LL}_{12}$ with the ratio from $10^{-3}$ to $10^{-7}$ as $m_{W_{R}}$
varies from 1 to 15 TeV.  Nevertheless, there are four terms which
appear in $M_{12}^{LR}$ in the case of st-coupling, namely those from $ 
(c,c)$, $(c,t)$, $(t,c)$ and $(t,t)$ pairs, and which are dominated by the
$(t,t)$ pair.  This gives $M_{12}^{LR} \sim M_{12}^{LL}$ for
$m_{W_{R}} = 2.5$ TeV, and $M^{LR}_{12} \ll M^{LL}_{12}$ by two orders
of magnitude for $m_{W_{R}} = 10 $ TeV.  The $W_{R}$ contribution to
$B_{s}-{\bar B}_{s}$ mixing cannot be ignored since $m_{W_{R}} < 5$ TeV in 
this case.
 \subsubsection{dt-coupling}
There is also one non-vanishing term in $M_{12}^{LR}$ coming from the
$(c,u)$ pair in the case of dt-coupling, but $M^{LR}_{12} \ll M^{LL}_{12}$ by 
five orders of magnitude. Here, $W_{R}$ gives no contribution to
$B_{s}-{\bar B}_{s}$ mixing.

\section{Joint {\boldmath $\chi^{2}$} Analysis for CKM Matrix Elements}

We use six present experiments for the determination of the CKM
matrix elements angles $s_{23}$, $s_{13}$, and $\delta$. These are
results for the matrix elements $|V_{cb}|=0.0404\pm0.0016$ and 
$|V_{ub}/V_{cb}|=0.101\pm0.016$\cite{cassel}, for 
$\epsilon_{K}$ in the neutral $K$ system, for $B_{d}-{\bar B}_{d}$
mixing with $\Delta m_d=0.476 \pm 0.016~ {\rm ps}^{-1}$, for the probability
of each calculated $\Delta m_s$\cite{xs} ,
and $\sin{(2\beta)}=0.79 \pm 0.19$ from Belle, BaBar and CDF\cite{Belle}.
Since $W_{R}$ may also contribute to $b$-decay, we include the 
contraint on $V_{cb}$ as $|V_{cb}^{L}|^{2} + x_{LR} |V_{cb}^{R}|^{2}
= |V_{cb}|^{2}$, where the interfering term is neglected\cite{gronau-0}.
The $V_{ub}$ value gives a $b-d$ unitarity triangle side of length
$0.46 \pm 0.07$.  Since some discrepancies for new physics will be very
large, our conclusions are not dependent on which choice of $V_{cb}$ or
$V_{ub}$ experiments are used.


For making projected experimental plots 
for pairs of experiments $(\sin{(2\alpha)}, \sin{(2\beta)})$, 
$(x_s,\sin{(\gamma)})$, or $(x_s,\sin{(2\phi_s)})$,
we add one of these pairs as two future experiments,
and assign as their errors the bin widths, which are 5\% of the total
range in our $20 \times 20$ bin coverage.  For 
$(\sin{(2\alpha)}, \sin{(2\beta)})$,
these errors are close to those achievable for the
$B$ factories.  Counting degrees of freedom, we
have for the case of st-coupling:  df = 8 experiments - 3 SM angles - 2 LR angles 
= 3 
df. For the case of dt-coupling we have: df = 8 experiments - 3 SM angles - 1 
LR angles = 4 df.

We produce the maximum likehood correlation plots for
$(\sin{(2\alpha)}, \sin{(2\beta))}$, $(x_{s},
\sin{(\gamma))}$, and $(x_s, A_{B_s})$.  
For each possible bin with given values for these
pairs, we search for the lowest $\chi^2$ in the data sets of the four
or five angles of $V^{L}$ and $V^{R}$, depending upon which case in
$V^{R}$ we are dealing with.  We then draw contours at a few values of
$\chi^{2}$ in these plots corresponding to given confidence
levels\cite{silverman} which match 1$\sigma$ and 2$\sigma$ limits.

\section{{\boldmath $CP$} Asymmetries in {\boldmath $B^0$} Decays}

\subsection{{\boldmath $(\sin{(2\alpha)},\sin{(2\beta)})$} Plots}
The first $CP$ violating asymmetry in $B \rightarrow J/\psi K_{S}$ decays is 
 related to the mixing matrix element $M_{12}$ and the decay amplitudes as 
follows\cite{rev},
\begin{equation} 
\sin{(2\beta)} \equiv - {\rm Im} \left( { M^{*}_{12} \over |M_{12}| } 
{ A(\bar{B} \rightarrow \Psi K_{s}) \over A(B \rightarrow \Psi K_{s}) } 
\right).   
\end{equation}                                  
The second $CP$ asymmetry
is provided by the measurement of the asymmetry in $B \rightarrow \pi 
\pi$, namely\cite{rev} 
\begin{equation}                      
\sin{(2\alpha)} \equiv {\rm Im} \left( { M^{*}_{12} \over |M_{12}| } 
{ A(\bar{B} \rightarrow \pi \pi) \over A( B  \rightarrow \pi \pi) } \right).   
\end{equation}

Because of the non-SM contributions of the LRSM, the effective
$\alpha$ and $\beta$
as defined here no longer represent real angles in the unitarity triangle.

\subsubsection{ st-coupling} 
Penguin diagrams, dominated by internal 
top-loops, also 
contribute to  $B \rightarrow J/\psi K_{S}$ decays 
in addition to the tree graphs. 
The phase of the $W_{R}$ penguin amplitude,
$V_{tb}^{R}V^{R*}_{ts}=cse^{i(\chi-\phi)}$, has
exactly the same value but opposite sign to
the phase of the $W_{R}$ tree amplitude, 
$V_{cb}^{R}V_{cs}^{R*} = cse^{i(\sigma-\xi)}$, 
due to the unitarity 
condition on $V_I^R$: $\chi-\phi=\sigma-\xi + \pi$.
Namely, $V_{tb}^{R}V_{ts}^{R*} = 
- V_{cb}^{R}V_{cs}^{R*}.$
 In the SM, the phase of  the 
$W_L$ penguin amplitude induced by an internal 
top-loop is also opposite to the phase of the 
$W_{L}$ tree amplitude, since 
$V_{tb}^{L}V_{ts}^{L*} \simeq - A\lambda^{2} 
\simeq - V_{cb}^{L}V_{cs}^{L*}.$ Consequently, recalling that 
$x_{LR} = m^2_{W_L}/m^2_{W_R}$, 
\begin{equation}
{ A(\bar{B} \rightarrow J/\psi K_S) \over A(B\rightarrow J/\psi K_S) }
= { V_{cb}^{L}V_{cs}^{L*}(1-P)/m_{W_{L}}^{2} +
V_{cb}^{R}V_{cs}^{R*}(1-P')/m_{W_{R}}^{2} 
\over  V_{cb}^{L*}V_{cs}^{L}(1-P)/m_{W_{L}}^{2} +
V_{cb}^{R*}V_{cs}^{R}(1-P')/m_{W_{R}}^{2} }
= { V_{cb}^{L}V_{cs}^{L*} + x_{LR}
V_{cb}^{R}V_{cs}^{R*}
\over  V_{cb}^{L*}V_{cs}^{L} + x_{LR}
V_{cb}^{R*}V_{cs}^{R} },
\end{equation}
where $P$ and $P'$ are the ratios of the $W_{L}$ and $W_{R}$ penguin 
contributions over the tree amplitudes, respectively, and
the approximation \cite{kurimoto} $P \cong  P'
\propto \alpha_{s}{\rm ln}(m_{t}^{2}/m_{c}^{2})$ provides
the simplification of $P = P'$. This gives
\begin{equation}
\sin(2\beta) = -{\rm Im}\left( {M^{*}_{12} \over |M_{12}|} { 
V_{cb}^{L}V_{cs}^{L*} + x_{LR} V_{cb}^{R}V_{cs}^{R*}
\over  V_{cb}^{L*}V_{cs}^{L} + x_{LR}
V_{cb}^{R*}V_{cs}^{R} }\right).
\end{equation}
On the other hand, no tree or penguin $W_{R}$ contributions exist in 
$B \rightarrow \pi \pi$ 
since $V_{ub}^{R}V_{ud}^{R*}=0$ and $V_{tb}^{R}V_{td}^{R*}=0$.
There is a $\Delta I = 1/2$ penguin pollution for this decay 
mode in the SM, but it can be removed by isospin analysis\cite{gronau}.  
Therefore, we have
\begin{equation}                     
\sin{(2\alpha)} = {\rm Im} \left({M^{*}_{12} \over |M_{12}|} 
{ V_{ud}^{L*}V_{ub}^{L}  \over V_{ud}^{L}V_{ub}^{L*} } \right).
\end{equation}
Since $M^{LR}_{12} \ll M^{LL}_{12}$ for $B_{d}$ mesons, the result is that
$M_{12} \simeq M^{LL}_{12}$. For  $m_{W_R} = 10$ TeV, with little $W_R$ effect, 
the ranges at 1$\sigma$ are $0.6 \le \sin{2(\beta)} \le 0.8$ and 
$-0.4 \le \sin{(2\alpha)} \le 0.1$.  However, for $m_{W_R} = 1.0$ TeV, the
ranges are $-0.5 \le \sin{(2\alpha)} \le 0.4$ and $0.50 \le \sin{(2\beta)} \le 0.95$.
\subsubsection{dt-coupling}
There are no tree or penguin contributions
by $W_{R}$ in $B \rightarrow J/\psi K_{S}$ due to the fact 
that $V_{cb}^{R}V_{cs}^{R*}=0$ and $V_{tb}^{R}V_{ts}^{R*}=0$.
Hence, one has 
\begin{equation}                     
\sin(2\beta) = -{\rm Im}\left( {M^{*}_{12} \over |M_{12}|} { 
V_{cb}^{L}V_{cs}^{L*} \over  V_{cb}^{L*}V_{cs}^{L} }\right).
\end{equation}
In the case of dt-coupling, Eq.~(28) still holds for the same 
reason, namely that the tree and penguin $W_{R}$ contributions 
do not exist in this case. 
 Fig.~1 shows the $(\sin{(2\alpha)},\sin{(2\beta))}$ plots
for the LRSM for values of $m_{W_{R}} =$ 2.5, 5, 7.5 and 10 TeV,
respectively, with contours at $\chi^{2}$ which correspond to
confidence levels for $1\sigma$ and $2\sigma$ limits.
We do not include the plot for $m_{W_{R}}=1$ TeV because 
$1\sigma$
and $2\sigma$ contours do not appear in the graph with such a low
value of $m_{W_{R}}$. The 
contributions for $m_{W_R} < 7.5$ TeV are very
different from those in the SM since $M^{LR}_{12} \simeq 
M^{LL}_{12}$ in this case.  The contours at $m_{W_R} = 10$ TeV 
should not
be directly compared with SM fit contours, which are smaller,
since the $W_R$ has ``decoupled'' 
here along with its two angles.  For the SM fits the 
$df = 8 - 3 = 5$ rather
than the $df = 3$ used for the plots here when $W_R$ is 
effective.

\subsection{{\boldmath $(x_s,\sin{(\gamma)})$} Plots}
The third asymmetry angle in $B$ meson systems is defined from
$B_s \to D_s^{+} K^{-}$ decays as\cite{aleksan}
\begin{equation}
\sin{(\gamma)} \equiv
{\rm Im}\left({ M^{B_s}_{12} \over |M_{12}^{B_s}| } 
{  A(B_{s} \rightarrow D^{+}_{s}K^{-}) \over 
   A(\bar{B}_{s} \rightarrow D^{+}_{s}K^{-}) }\right).
\end{equation}
The penguin contribution is absent in both 
$B_{s} \rightarrow D^{+}_{s}K^{-}$ and  
$\bar{B}_{s} \rightarrow D^{+}_{s}K^{-}$ decays.
Again, because of the LRSM contribution, $\gamma$ as defined 
above is no longer an angle of the unitarity triangle.
$x_{s}$ is given here by 
\begin{equation}
x_{s} = 1.034 x_{d} {|M^{B_{s}}_{12}| \over |M_{12}^{B}|}. 
\end{equation}
\subsubsection{st-coupling}
The contributions from
$W_{R}$ to both decay modes vanish since
$V_{ub}^{R*}V_{cs}^{R}=0$ and $V_{cb}^{R}V_{us}^{R*}=0.$ Therefore,
the $CP$ asymmetry for this decay mode can be simplified as 
\begin{equation}
\sin{(\gamma)} =
{\rm Im}\left({M^{B_{s}}_{12} \over |M_{12}^{B_s}|} 
{V_{ub}^{L*}V_{cs}^{L} \over V_{cb}^{L}V_{us}^{L*}}
/\left|{V_{ub}^{L*}V_{cs}^{L} \over V_{cb}^{L}V_{us}^{L*}}\right|
\right). 
\end{equation}
The $(x_{s}, \sin{(\gamma))}$ 
plots of the case of st-coupling are shown in Fig. 2.  
In the SM, with the same parameters that we used, 
$\sin{(\gamma)}$ has a range $0.6 \le \sin{(\gamma)} \le 0.9$
at 1$\sigma$, but in the
LRSM at low $m_{W_R}$, $\sin{(\gamma)}$ can extend completely from
-1 to 1 at 1 TeV, and from 0 to 1 at 2.5 TeV.
Comparing to the
range of $x_{s}$ in the SM (with the parameters in this paper), which is from 
$20$ to $40$ at 
$1\sigma$, $x_{s}$ has a range of about 20 to 50 
for $m_{W_{R}} = 2.5$ TeV and greater than 100 for $m_{W_{R}}=1.0$ TeV.  
This is because
$M_{12}^{B_{s}}$ is almost double that in the SM, 
while $M_{12}^{B}$
behaves similarly to that of the SM.  This amplification is 
then reduced
as $m_{W_{R}} \sim 5$ TeV, and finally the ratio in $x_s$, Eq. (31), 
approaches the SM result.  
\subsubsection{dt-coupling}
Because $W_{R}$ can contribute to $\bar{B}_{s} \rightarrow D_{s}^{+}K^{-}$,
in this case we have
\begin{equation}
\sin{(\gamma)} = 
{\rm Im}\left({M^{B_{s}}_{12} \over |M_{12}^{B_s}|} 
{V_{ub}^{L*}V_{cs}^{L} \over 
 V_{cb}^{L}V_{us}^{L*} +
 x_{LR} V_{cb}^{R}V_{us}^{R*}}
/\left|{V_{ub}^{L*}V_{cs}^{L} \over V_{cb}^{L}V_{us}^{L*} +
 x_{LR} V_{cb}^{R}V_{us}^{R*}} \right|
\right).  
\end{equation}
In the $(x_{s}, \sin{(\gamma))}$ plot for the case of dt-coupling,
 $x_{s}$ has about the same range as in the SM for $m_{W_{R}} \ge 2.5$  
TeV.  
\subsection{{\boldmath $(x_s, \sin{(2\phi_s)})$} Plots}
The asymmetry $\sin{(2\phi_s)}$ for $B_s-\bar{B}_s$ mixing is 
given by 
\begin{equation}
\sin{(2\phi_s)} \equiv - {\rm Im}\left({M^{B_{s}}_{12} \over |M_{12}^{B_s}|} 
{A(\bar{b} \rightarrow \bar{c}c\bar{s}) \over A(b \rightarrow  
c\bar{c}s) }\right), 
\end{equation} 
where $\phi_s$ is also the small angle in the $b-s$ unitarity triangle in
the SM.  In the standard model,
$\sin{(2\phi_s)}$ is almost zero $(\approx 0.025)$ in
due to the fact that
neither the decay process of ${\bar b} \rightarrow 
\bar{c}c\bar{s}$ nor the mixing effect in $B_{s}$ 
provides much phase to the asymmetry.
\subsubsection{st-coupling}
Both $W_{L}$ and $W_{R}$ can contribute to ${\bar b} \rightarrow 
\bar{c}c{\bar s}$ in this case. This implies 
\begin{equation}
\sin{(2\phi_s)} = -{\rm Im} \left( {M^{B_{s}}_{12} \over |M_{12}^{B_{s}}|} 
{V_{cb}^{L*}V_{cs}^{L} + x_{LR} V_{cb}^{R*}V_{cs}^{R}
\over  V_{cb}^{L}V_{cs}^{L*} + x_{LR}
V_{cb}^{R}V_{cs}^{R*} }\right).
\end{equation}
$M_{12} \simeq
M_{12}^{LL} + M_{12}^{LR} + M_{12}^{RR} \simeq 
M_{12}^{LL} + M_{12}^{LR}$ with $M_{12}^{LL}
\simeq M_{12}^{LR}$ as $m_{W_{R}} \le 2.5$ TeV  for 
$B_{s}$ mesons. 
$M_{12}^{LR}$ is dominated by the $(t,t)$ pair as shown
in Eq. (22), and this term can provide a non-vanishing phase to 
the asymmetry $\sin{(2\phi_s)}$. In this case, $\phi_s$ is no longer
an angle in a unitarity triangle, although the measured asymmetry will
be called $\sin{(2\phi_s)}$.
The  $(x_{s}, \sin{(2\phi_s)})$ plots for the case of st-coupling
are shown in Fig. 3. $\sin{(2\phi_s)}$ can be zero or maximal at $\pm 1$ at the
$1\sigma$ level for $m_{W_R} \le 2.5$ TeV, and very large even for $5.0$ TeV. 
This distinction from the small SM result at $m_{W_R} = 10$ TeV 
can provide a dramatic and clean test of
new physics. 
\subsubsection{dt-coupling}
There is no $W_{R}$ contribution in $\bar{b} \rightarrow {\bar c}c\bar{s}$ 
decays. Thus,
\begin{equation}
\sin{(2\phi_s)} = - {\rm Im}\left( { M^{B_{s}}_{12} \over |M_{12}^{B_s}| } 
{ V_{cb}^{L*}V_{cs}^{L} \over V_{cb}^{L}V_{cs}^{L*} } \right).
\end{equation} 
The fact that $ M_{12}^{LR} < 10^{-3} M_{12}^{LL}$ makes 
 $M_{12} \simeq
M_{12}^{LL}$ for the $B_{s}$ system.
Hence, the asymmetry $\sin{(2\phi_s)}$ is almost zero
and the same as that in the SM\cite{silverman}.
\subsection{{\boldmath $(\rho,\eta)$} Plots}
We define the $(\rho,\eta)$ point from the $V^L$ Wolfenstein
form as 
\begin{equation}
\rho + i\eta = V^{L*}_{ub}/|V^{L*}_{cb} V^L_{cd}|.
\end{equation}
\subsubsection{st-coupling}
This case is explained in subsection A, where for $m_{W_R} \ge
1.0$ TeV, $M_{12} \simeq M_{12}^{LL}$, and the plots are about
the same as in the SM and are independent of $m_{W_R}$.
\subsubsection{dt-coupling}
Fig. 4 shows the $(\rho,\eta)$ plots for $m_{W_R} = 2.5, 5.0, 
7.5$ and $10.0$ TeV.  For the lowest value, $m_{W_R} = 2.5$ TeV,
we see in addition to the SM oval, a second oval region 
to the left of the SM region at 2$\sigma$.
\section{Conclusions}
In order to provide
a reasonable lower limit for the $W_{R}$ mass within the 
$SU(2)_{L} \times 
SU(2)_{R} \times U(1)$ model, the right handed quark 
mixing matrices can be parametrized into two forms or 
cases\cite{langacker}
as shown in Eq. (3). We suppress the large contributions to 
$\epsilon_{K}$
from the $W_{L} - W_{R}$ box diagram by effectively taking some 
parameters of
$V^{R}$ to vanish, as depicted in Eqs. (14) and (15), so that 
the quite small
experimental value of $\epsilon_{K}$ can be satisfied and $W_{R}$
can give the most significant effects on $CP$ asymmetries in $B$ 
decays\cite{kurimoto}.

The LRSM can contribute importantly to $B_{d} - \bar{B}_{d}$ 
mixing  
in the case of dt-coupling and to $B_{s} - \bar{B}_{s}$ mixing  in the case 
of st-coupling, and give new phases, for $m_{W_{R}} < 10$ TeV. Hence, $W_{R}$ 
shows its effects on $(\sin2{\alpha}, \sin2{\beta})$ in the case of           
dt-coupling
and on $(x_{s}, \sin{\gamma})$ in the case of st-coupling, for $m_{W_{R}} < 10$ 
TeV. The CP asymmetries in the LRSM for $m_{W_{R}} < 10$ TeV
that are different from those in the SM are:
(i) $\sin{(2\phi_s)}$ can be maximal at $\pm 1$ in the case of st-coupling; 
(ii) the range for $x_{s}$ is from
$20$ to $\ge 100$ at the $1\sigma$ level in the case of st-coupling; 
(iii) $\sin{\gamma}$ has a much larger range in
the case of st-coupling;
and (iv) $0.3 \le \sin{(2\alpha)} \le 1$ in the case of dt-coupling at 
$m_{W_R}=2.5$ TeV. 
If the experimental results are consistent with the SM at $1\sigma$ 
in
$(\sin{(2\alpha)}, \sin{(2\beta)})$ and in $(x_{s}, \sin{(\gamma)})$, 
the LRSM cannot
be ruled out, but the limit $m_{W_{R}} \ge 10$ TeV will be 
established.  What is
striking is that the asymmetry $\sin{(2\phi_s)}$ for 
$B_{s} - \bar{B}_{s}$
mixing is clearly far from zero at the $1\sigma$ level even for
$m_{W_{R}} \simeq 10$ TeV in the case of st-coupling, as shown in Fig. 3. This 
difference from
the SM can provide a clean test of new physics.

\section*{Acknowledgments}
D. S. is supported in part by the 
U. S. Department of Energy under Contract No. DE-FG0391ER40679.

\begin{figure}

\vspace{3mm}

\caption{ The $(\sin{(2\alpha)}, \sin{(2\beta))}$ plots for the 
left-right symmetric model in the case of dt-coupling for values of 
(a) $m_{W_{R}}=2.5$, (b)
  $m_{W_{R}}=5$, (c) $m_{W_{R}}=7.5$, and (d) $m_{W_{R}}=10$ TeV.
  Contours are at $1\sigma$ and $2\sigma$.}
\label{ab2}
\end{figure}

\begin{figure}
\caption{ The $(x_{s}, \sin{(\gamma))}$ plots for the 
left-right symmetric
  model in the case of st-coupling for values of (a) $m_{W_{R}}=1.0$,
(b) $m_{W_{R}}=2.5$, (c) $m_{W_{R}}=5.0$, and  (d) $m_{W_{R}}=10$ 
TeV, with contours at $1\sigma$ and $2\sigma$.}
\label{xs1}
\end{figure}

\begin{figure}
\caption{ The $(x_{s}, \sin{(2\phi_s)})$ plots for the $B_{s}$ 
asymmetry $\sin{(2\phi_s)}$ in the left-right symmetric
  model in the case of st-coupling for values of (a) $m_{W_{R}}=1.0$,
(b) $m_{W_{R}}=2.5$, (c) $m_{W_{R}}=5.0$, and  (d) $m_{W_{R}}=10$ 
TeV.  Contours are at $1\sigma$ and $2\sigma$.}
\label{bs1}
\end{figure}

\begin{figure}
\caption{ The $(\rho, \eta)$ plots for the left-right symmetric
  model in the case of dt-coupling for values of (a) $m_{W_{R}}=2.5$,
(b) $m_{W_{R}}=5$, (c) $m_{W_{R}}=7.5$, and  (d) $m_{W_{R}}=10$ 
TeV, with contours at $1\sigma$ and $2\sigma$.}
\label{re2}
\end{figure}

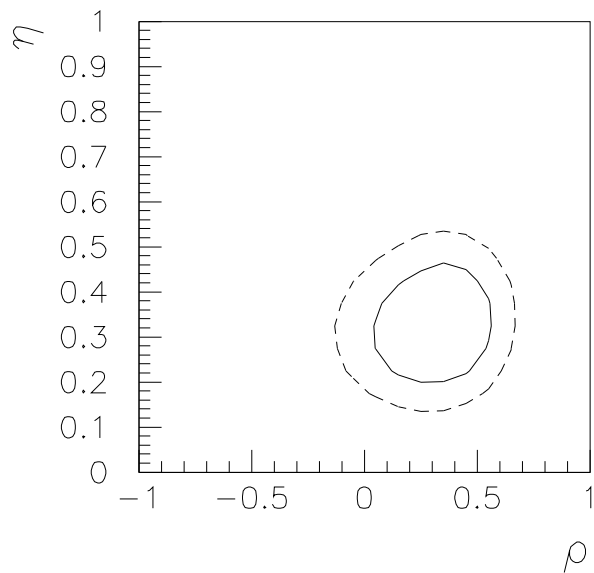
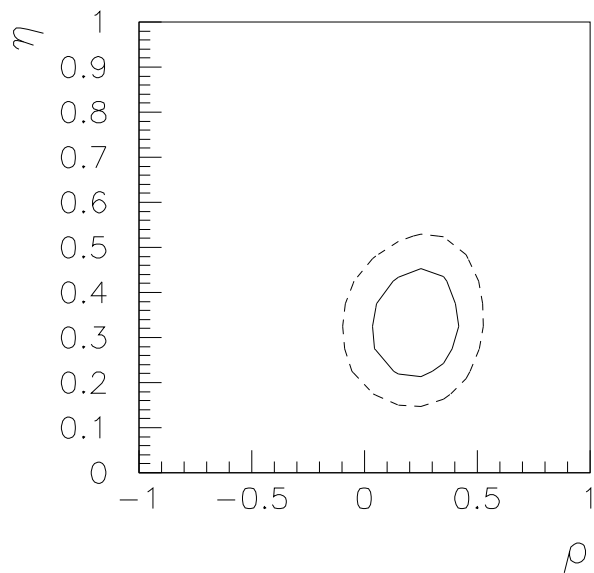
\end{document}